\documentclass[12pt]{article}

\usepackage{amssymb}
\usepackage{amsthm}
\newtheorem {theorem} {Theorem}

\newtheorem {corollary} [theorem]{Corollary}

\newtheorem {remark} [theorem]{\bf Remark}
\newtheorem {example} [theorem]{Example}

\newcommand{\bbox}{\ \hfill\rule[-1mm]{2mm}{3.2mm}}

\title{Perturbed Euler top and bifurcation of limit cycles on invariant Casimir surfaces}

\author{Isaac A. Garc\'{\i}a$^{\ 1,*}$ and Benito Hern\'andez-Bermejo$^{\ 2}$}

\date{$^{\ (1)}$ {\small Departament de Matem\`atica. Universitat de
Lleida. \\ Avda. Jaume II, 69. 25001 Lleida, Spain.
\\ E--mail: {\tt garcia@matematica.udl.cat}
\\ $ $ \\
$^{\ (2)}$ Departamento de F\'{\i}sica. Universidad Rey Juan Carlos.
\\ Calle Tulip\'{a}n S/N. 28933--M\'{o}stoles--Madrid, Spain.
\\ E-mail: {\tt benito.hernandez@urjc.es}}}

\begin{document}

\maketitle

\begin{abstract}
Analytical perturbations of the Euler top are considered. The perturbations are based on
the Poisson structure for such a dynamical system, in such a way that the Casimir
invariants of the system remain invariant for the perturbed flow. By means of the
Poincar\'{e}-Pontryagin theory, the existence of limit cycles on the invariant Casimir
surfaces for the perturbed system is investigated up to first order of perturbation,
providing sharp bounds for their number. Examples are given.
\end{abstract}

\noindent {\bf Keywords:} Poisson systems; Casimir invariants; Hamiltonian systems; perturbation theory; limit cycles; Poincar\'{e}-Pontryagin theory.

\mbox{}

\mbox{}

\noindent {\bf PACS codes:} 02.30.Hq, 05.45.-a, 45.20.-d, 45.20.Jj.


\vfill

\footnoterule

$^*$ Corresponding author. Telephone: (+34) 973702728. Fax: (+34) 973702702.

\pagebreak
\section{The Euler top}

The basis of this work is the following system of ODEs known as Euler equations, which describe the rotation of a rigid body, or Euler top:
\begin{equation}\label{top}
    \left\{ \begin{array}{ccl}
    \dot{x}_1 & = & \displaystyle{ \frac{\mu _2 - \mu _3}{\mu _2 \mu _3}x_2x_3 } \ ,
        \vspace{3mm} \\
    \dot{x}_2 & = & \displaystyle{ \frac{\mu _3 - \mu _1}{\mu _3 \mu _1}x_3x_1 } \ ,
        \vspace{3mm} \\
    \dot{x}_3 & = & \displaystyle{ \frac{\mu _1 - \mu _2}{\mu _1 \mu _2}x_1x_2 } \ .
    \end{array} \right.
\end{equation}
Here ${\bf x} = (x_1, x_2, x_3) \in \mathbb{R}^3$ where $x_i$ denotes the $i$th component of angular momentum, and constants $\mu _i$ are the moments of inertia about the coordinate axes, both for $i=1,2,3$. Energy is conserved for this system, but of course the flow is odd dimensional and a classical Hamiltonian formulation is excluded. However this is a Poisson system (see \cite{lic1,olv1,wei1} for general references on Poisson systems) in terms of the following structure matrix:
\[
     {\cal J}({\bf x}) = \left( \begin{array}{ccc}
                    0  & -x_3 & x_2  \\
                   x_3 &  0   & -x_1 \\
                   -x_2 & x_1  &  0
              \end{array} \right) \ .
\]
Notice that the rank of the structure matrix is 2 everywhere in $\mathbb{R}^3$ except at the origin, in which the rank vanishes. The Hamiltonian is the total energy (kinetic energy, in this case):
\[
    \mathcal{H}(\mathbf{x}) = \frac{1}{2} \left( \frac{x_1^2}{\mu _1} + \frac{x_2^2}{\mu _2} +
    \frac{x_3^2}{\mu _3} \right) \ .
\]
This $SO(3)$-based Lie-Poisson structure of the Euler top equations was first recognized by Arnol'd \cite{arn1} (see \cite{tym1} for a modern classification of Lie-Poisson structures).  In addition, other Poisson formulations are also possible for system (\ref{top}), for instance see \cite{gyn1,bs4,bs5}, although we shall not be concerned with them in this work. Since its characterization, the $SO(3)$ Lie-Poisson structure for the Euler top has been repeatedly investigated from a variety of perspectives and has provided the basis for a number of mathematical and physical developments, e.g. see \cite{agz1,creh,gmp1,gyn1,bs2,hm1,pkrm,put1,sim1}.

Since the rank of ${\cal J}$ is 2 (excluding the origin from the analysis) there must be one independent Casimir invariant, which can be chosen to be:
\begin{equation}
\label{nmaetop}
    D(\mathbf{x}) = x_1^2+x_2^2+x_3^2 = \| \mathbf{x} \|^2 \ .
\end{equation}
Namely, the distinguished invariant (\ref{nmaetop}) is the square of the Euclidean norm of the angular momentum, which is a conserved quantity during the system rotation. Therefore, the symplectic foliation is given by:
\begin{equation}
\label{slaetop}
    x_1^2+x_2^2+x_3^2 = \mbox{\rm constant}
\end{equation}
which are concentric spheres in $\mathbb{R}^3$. We thus see that the symplectic leaves
(\ref{slaetop}) are even dimensional (two dimensional, in this case) and therefore Darboux's theorem ensures that on the symplectic leaves the dynamics is Hamiltonian in the classical sense, at least locally (in the neighborhood of each point). The actual trajectories of the system in phase space are obtained by the intersection of the symplectic leaves with the energy level sets
\[
    \mathcal{H}(\mathbf{x}) = \frac{1}{2} \left( \frac{x_1^2}{\mu _1} + \frac{x_2^2}{\mu _2} +
    \frac{x_3^2}{\mu _3} \right)
    = \mbox{\rm constant}
\]
which in geometric terms are ellipsoids in phase space.

\section{Bifurcation of limit cycles after perturbation of planar Hamiltonian centers}

Oscillations play a prominent role in many physical systems where an important problem is to determine if spontaneous oscillatory activity persists when subjected to a small external stimulus. In this sense, it is worth recalling here the relationship between the number of zeros of the so-called Poincar\'e--Pontryagin function given by a generalization of an Abelian integral and the number of limit cycles (isolated periodic solutions) of the corresponding planar analytic differential systems.

We consider an analytic Hamiltonian function $H(x, y)$ defined on some open subset $U \subseteq \mathbb{R}^2$. We shall assume that, the corresponding Hamiltonian vector field $\mathcal{X}_H$ has a family of periodic orbits filling up an annulus and
given by the ovals $\gamma_h \subset H^{-1}(h)$, continuously depending on a parameter $h \in (a, b) \subset \mathbb{R}$. Now, we perturb the system as follows
\begin{equation} \label{Euler-Pontryagin1}
\dot{x} = \frac{\partial H}{\partial y} + \epsilon P(x, y) \ , \ \dot{y}  = - \frac{\partial H}{\partial x} + \epsilon Q(x, y) \ ,
\end{equation}
where $P$ and $Q$ are analytic functions on $U$ and $\epsilon$ is a real small parameter. We emphasize here to the reader that the considered
perturbations (\ref{Euler-Pontryagin1}) are not necessary Hamiltonian ones.

Then, we may define the Poincar\'e--Pontryagin function $I(h)$ as the following line integral
\begin{equation} \label{Euler-Pontryagin2}
I(h) = \oint_{\gamma_h} P(x, y) dy - Q(x, y) dx \ .
\end{equation}
Notice that, in case that $H$, $P$ and $Q$ be polynomials, then $I(h)$ is just an Abelian integral. For small values of $\epsilon$, the following question arises: How many orbits $\gamma_h$ keep unbroken and become periodic orbits of the perturbed system (\ref{Euler-Pontryagin1})? Clearly, if this number is finite, then they are limit cycles of (\ref{Euler-Pontryagin1}). We say that a periodic orbit $\Gamma_\epsilon$ of the perturbed system (\ref{Euler-Pontryagin1}) bifurcates from $\gamma_h$ if $\Gamma_\epsilon$ tends to $\gamma_h$ in the sense of Hausdorff distance as $\epsilon \to 0$.

In the generic case $I(h) \not\equiv 0$, the answer of the aforementioned question is given by the next theorem, see for instance \cite{Ch-Li} for its proof and the state--of--the--art of this and other related topics.

\begin{theorem}\label{Euler-Pontryagin3}
Let $I(h)$ be given by (\ref{Euler-Pontryagin2}). Assuming $I(h) \not\equiv 0$ for $h \in (a, b)$, the following statements hold:
\begin{description}
\item[(i)] If system (\ref{Euler-Pontryagin1}) has a limit cycle bifurcating from $\gamma_{h^*}$, then $I(h^*) = 0$.

\item[(ii)] If $I(h)$ has a simple zero at $h^* \in (a, b)$ (that is, $I(h^*) = 0$ and $I'(h^*) \neq 0$), then system (\ref{Euler-Pontryagin1})
has a unique limit cycle bifurcating from $\gamma_{h^*}$ and, moreover, this limit cycle is hyperbolic.
\end{description}
\end{theorem}

In this work, analytical perturbations of the Euler top (\ref{top}) shall be investigated, in such a way that either some invariant surface $\mathbb{S}_c = \{ (x_1, x_2, x_3) \in \mathbb{R}^3 \ : \ D(x_1, x_2, x_3) = c^2 \}$ for an arbitrary fixed $c>0$ remains invariant or all the invariant surfaces $\mathbb{S}_c$ for any $c>0$ remain invariant. Actually, such kind of analysis is natural because Casimir-preserving perturbations of the Euler top have been reported in the literature, in the context of stability and control theory \cite{Puta,put1}. After that, we will study up to first order the existence of limit cycles on $\mathbb{S}_c$ for the perturbed system by means of Theorem \ref{Euler-Pontryagin3}. To conclude, some examples are given.

Let us recall here that, when the phase space has dimension greater than 2, a periodic
orbit $\gamma$ is called {\it limit cycle} if it is $\alpha$ or $\omega$--limit set of
another orbit. Thus, in this case $\gamma$ needs not to be isolated inside the set of
periodic orbits. Several authors have studied (by using different techniques)
bifurcations of limit cycles in perturbations of a vector field with an invariant
two--dimensional manifold, assuming that the restriction of the field on this manifold is
Hamiltonian (see for instance \cite{Ci-Lli-Te,Zol1}). To the authors' knowledge, this
kind of analysis is carried out for the Casimir invariants of a Poisson system for the
first time in the present work. Moreover, the preservation of the Casimir invariants is
physically relevant, as far as it amounts to a conservation of the angular momentum for
the perturbed system. By suppressing the explicit restriction about the existence of
invariant two--dimensional manifolds, in \cite{Bu-Ga} and using Poincar\'e maps, a study of
the existence of $T$--periodic solutions for $T$--periodic perturbations of the symmetric
($\mu_1=\mu_2$) Euler top is performed.

\section{Perturbed Euler top}

Let us rewrite the Euler top (\ref{top}) as
\begin{equation} \label{Euler1}
\dot{x}_1  = \alpha x_2 x_3 \ , \ \dot{x}_2  = \beta x_1 x_3 \ , \dot{x}_3  = \gamma x_1 x_2 \ ,
\end{equation}
with parameters
$$
\alpha := \frac{\mu_2-\mu_3}{\mu_2 \mu_3} \ , \ \beta := \frac{\mu_3-\mu_1}{\mu_1 \mu_3} \ , \ \gamma := \frac{\mu_1-\mu_2}{\mu_1 \mu_2} \ .
$$
Notice that $\alpha+\beta+\gamma=0$. The Euler top is an integrable system having the first integrals
$$
\mathcal{H}(x_1,x_2,x_3) = \frac{1}{2} \left( \frac{x_1^2}{\mu_1}+ \frac{x_2^2}{\mu_2}+ \frac{x_3^2}{\mu_3} \right) \ , \ D(x_1,x_2,x_3) = x_1^2+x_2^2+x_3^2 \ .
$$
Without loss of generality, we can assume the condition $\alpha \beta < 0$. Therefore, given any real value $c \neq 0$, the invariant spheres
\begin{equation} \label{Euler1-sphere}
\mathbb{S}_c := \{ (x_1, x_2, x_3) \in \mathbb{R}^3 \ : \ D(x_1, x_2, x_3) = c^2 \} \ ,
\end{equation}
are foliated by periodic orbits of (\ref{Euler1}). We will also define the semispheres
\begin{equation} \label{Euler1-sphere2}
\mathbb{S}_c^{+} :=  \{ (x_1, x_2, x_3) \in \mathbb{S}_c \ : \ x_3 > 0 \} \ , \ \mathbb{S}_c^{-} :=  \{ (x_1, x_2, x_3) \in \mathbb{S}_c \ : \ x_3 < 0 \} \ .
\end{equation}
Now, we consider the following analytic perturbation in $\mathbb{R}^3 \backslash \{ x_3 = 0 \}$ of the Euler top (\ref{Euler1}) leaving invariant the semispheres $\mathbb{S}_c^{+}$ and $\mathbb{S}_c^{-}$.
\begin{eqnarray}
\dot{x}_1  &=& \alpha x_2 x_3 + \epsilon A(x_1,x_2,x_3) \ , \nonumber \\
\dot{x}_2  &=& \beta x_1 x_3 + \epsilon B(x_1,x_2,x_3)  \ , \label{Euler9} \\
\dot{x}_3  &=& \gamma x_1 x_2 + \epsilon C(x_1,x_2,x_3) \ , \nonumber
\end{eqnarray}
where
\begin{eqnarray}
A(x_1,x_2,x_3)  &=& x_3 P(x_1,x_2, D(x_1,x_2,x_3)) \ , \nonumber \\
B(x_1,x_2,x_3)  &=& x_3 Q(x_1,x_2, D(x_1,x_2,x_3))  \ , \label{condition-banach}\\
C(x_1,x_2,x_3)  &=& \frac{D(x_1,x_2,x_3)-c^2}{2 x_3}  R(x_1,x_2, D(x_1,x_2,x_3)) \nonumber \\
 & & -x_1 P(x_1,x_2, D(x_1,x_2,x_3)) - x_2 Q(x_1,x_2, D(x_1,x_2,x_3)) \ , \nonumber
\end{eqnarray}
being $P$, $Q$ and $R$ analytic functions in all $\mathbb{R}^3$. The following theorem is one of the main results of this work. Without loss of generality, in statements (ii) and (iii) of the theorem we shall focus on $\mathbb{S}_c^{+}$.

\begin{theorem}\label{teoeuler1}
Let us consider the Euler top (\ref{Euler1}) with $\alpha \beta < 0$, therefore having the invariant sphere $\mathbb{S}_c$ foliated with periodic orbits. Then, the following holds:
\begin{description}
\item[(i)] The most general analytic 1--parameter perturbation of (\ref{Euler1}) in $\mathbb{R}^3 \backslash \{ x_3 = 0 \}$ leaving invariant the semispheres $\mathbb{S}_c^{+}$ and $\mathbb{S}_c^{-}$ is given by (\ref{Euler9}--\ref{condition-banach}).

\item[(ii)] Assume that
\begin{equation} \label{Euler11-1}
I(h) = \oint_{H = h} P(x,y,c^2) dy - Q(x,y,c^2) dx \not\equiv 0 \ ,
\end{equation}
where $H(x,y) = \frac{1}{2} (\alpha y^2 - \beta x^2)$. Then, the periodic orbits $\gamma_{h^*} \subset \mathbb{S}_c^{+}$ of (\ref{Euler1}) from which bifurcates a limit cycle of the perturbed system (\ref{Euler9}) are given by
$$
\gamma_{h^*} = \{ (x_1, x_2, x_3) \in \mathbb{S}_c^{+} \ : \ \mathcal{H}(x_1, x_2, x_3) = c^2/(2 \mu_3) - h^* \} \ ,
$$
where $I(h^*)=0$, $\alpha \, h^* > 0$ and
$$
h^* < \frac{c^2}{2} \left( \frac{1}{\mu_3} - \max \left\{ -\alpha \: , \: \beta \right\} \right) \: , \:\;\: \mbox{or} \:\;\:
h^* > \frac{c^2}{2} \left( \frac{1}{\mu_3} - \min \left\{ -\alpha \: , \: \beta \right\} \right)
$$
if $\alpha > 0$ or $\alpha < 0$, respectively.

\item[(iii)] If $P$ and $Q$ are polynomials of maximum degree $n$ in the first two variables $x$ and $y$, then a sharp upper bound of the maximum number of limit cycles of system (\ref{Euler9}) on $\mathbb{S}_c^{+}$ is
\begin{equation}
\label{cotas}
\left\{ \begin{array}{lcl} (n-1)/2 & , & \mbox{if $n$ is odd} \\ (n-2)/2 & , & \mbox{if $n$ is even}.
\end{array} \right.
\end{equation}
\end{description}
\end{theorem}
\noindent{\bf Proof.} We perform the change of variables given by the diffeomorphism
\begin{equation}
\label{mistetas}
(x_1, x_2, x_3) \mapsto (x,y, z) \ , \ z=D(x_1, x_2, x_3) \ ,
\end{equation}
defined in $\{(x_1,x_2,x_3) \in \mathbb{R}^3 : \: x_3 > 0 \}$. Observe that under such transformation, the semisphere $\mathbb{S}_c^{+}$ is transformed into the open disk $\Omega = \{ (x,y, c^2) \in \mathbb{R}^3 : \:  x^2+y^2 < c^2 \}$. The perturbed system (\ref{Euler9}) restricted to the semispace $x_3 > 0$ adopts the form
\begin{eqnarray}
\dot{x} &=& \sqrt{z-(x^2+y^2)} \left( \frac{\partial H}{\partial y} + \epsilon P(x, y, z) \right) \ , \nonumber \\
\dot{y} &=& \sqrt{z-(x^2+y^2)} \left( -\frac{\partial H}{\partial x} + \epsilon Q(x, y, z) \right)  \ , \label{Euler12} \\
\dot{z} &=& \epsilon (z-c^2) R(x, y, z) \ , \nonumber
\end{eqnarray}
with $H(x,y) = \frac{1}{2} (\alpha y^2 - \beta x^2)$. The perturbed system (\ref{Euler9}) restricted to the semispace $x_3 < 0$ is given by (\ref{Euler12}) changing the sign in the right--hand side of $\dot{x}$ and $\dot{y}$. The expression of (\ref{Euler12}) contains the most general perturbation of the Euler top (\ref{Euler1}) written in $(x,y,z)$--coordinates which leaves invariant the disk $\Omega$. Therefore, undoing the change of coordinates done, statement (i) is proved.
\newline

We emphasize that, condition $\alpha \beta < 0$ implies that system (\ref{Euler12}) with  $\epsilon = 0$ possesses on the invariant disk $\Omega$ a center at $(x,y)=(0,0)$. Its period annulus is foliated by the ellipses $\gamma_h = H^{-1}(h)$ with certain values of $h$ to be specified later. Clearly, $h > 0$ when $\alpha > 0$ and $h < 0$ when $\alpha < 0$, if $\gamma_h$ is to be an ellipse. Moreover, since $\mathcal{H}(x, y, \sqrt{c^2-(x^2+y^2)}) = c^2/(2 \mu_3) - H(x,y)$, it follows that the level curve $\gamma_h$ corresponds to the level curve $\mathcal{H}(x, y, \sqrt{c^2-(x^2+y^2)}) = \bar{h}$ with $\bar{h} = c^2/(2 \mu_3) - h$. Finally, since  $\gamma_h \subset \Omega$, we must impose that both semiaxes $\sqrt{2 \bar{h}/\beta}$ and $\sqrt{-2 \bar{h}/\alpha}$ of $\gamma_h$ be smaller than the disk radius $c$ of $\Omega$. This last condition leads to two possibilities:
\begin{itemize}
\item If $\alpha >0$ and $\beta <0$, then it must be:
\[
   \bar{h} > \frac{c^2}{2} \max \left\{ - \alpha \: , \: \beta \right\}
\]
\item In the complementary case $\alpha <0$ and $\beta >0$, we have:
\[
    \bar{h} < \frac{c^2}{2} \min \left\{ - \alpha \: , \: \beta \right\} \ .
\]
\end{itemize}

Hence, the restriction of system (\ref{Euler12}) to the invariant disk $\Omega$ is given by the analytic system
\begin{eqnarray*}
\dot{x} &=& \sqrt{c^2-(x^2+y^2)} \left( \frac{\partial H}{\partial y} + \epsilon P(x, y, c^2) \right) \ , \\
\dot{y} &=& \sqrt{c^2-(x^2+y^2)} \left( -\frac{\partial H}{\partial x} + \epsilon Q(x, y, c^2) \right)  \ .
\end{eqnarray*}
Now, using Theorem \ref{Euler-Pontryagin3}, it is found that the periodic orbits of the center that persist under the perturbation for small $\epsilon$ are given by the zeros of (\ref{Euler11-1}). Then, statement (ii) is proved.
\newline

In order to compute (\ref{Euler11-1}), we will assume $\alpha > 0$ and $\beta < 0$ without loss of generality (otherwise, we change $t \to -t$). The ellipses $H = h$ with $h > 0$ can be parameterized as $x=\sqrt{-2 h /\beta} \cos\theta$, $y=\sqrt{2 h /\alpha} \sin\theta$ with $\theta \in [0,2 \pi)$. Then,
\begin{equation} \label{Euler8}
I(h) = \sqrt{h} \int_{0}^{2 \pi} ( \bar{P}(\theta, h, c) \cos\theta + \bar{Q}(\theta, h, c) \sin\theta )  d \theta \ ,
\end{equation}
where
\begin{eqnarray*}
\bar{P}(\theta, h, c) &:=& \sqrt{2/\alpha} \ P(\sqrt{-2 h / \beta} \cos\theta, \sqrt{2 h /\alpha} \sin\theta, c^2) \ , \\
\bar{Q}(\theta, h, c) &:=& \sqrt{-2/\beta} \ Q(\sqrt{-2 h /\beta} \cos\theta, \sqrt{2 h /\alpha} \sin\theta, c^2) \ .
\end{eqnarray*}
Assume now that the perturbation functions $P$ and $Q$ are polynomials in $x$ and $y$ of degree at most $n$, that is, $P(x,y,z) = \sum_{i+j=0}^n a_{ij}(z) x^i y^j$ and $Q(x,y,z) = \sum_{i+j=0}^n b_{ij}(z) x^i y^j$ with analytic coefficients $a_{ij}(z), b_{ij}(z) \in C^{\omega}(\mathbb{R})$.
After some algebra, we get that $I(h) = \sqrt{h} M_n(h)$ with $M_n(h)$ a polynomial in the variable $\sqrt{h}$ of maximum degree $n$. More precisely,
$$
M_n(h)= \sum_{i+j=0}^n \left[ \tilde{a}_{ij} I_{ij} + \tilde{b}_{ij} J_{ij} \right] (\sqrt{h})^{i+j}
$$
where $\tilde{a}_{ij}$ and $\tilde{b}_{ij}$ are real constants, and:
$$
I_{ij}= \int_{0}^{2 \pi} \cos^{i+1} \theta \sin^j \theta \: d \theta \: , \:\;\:\;
J_{ij}= \int_{0}^{2 \pi} \cos^i \theta \sin^{j+1} \theta \: d \theta \: .
$$
Taking into account that
\begin{equation}
\label{integrales}
\int_0^{2\pi} \sin^i \theta \cos^j \theta d \theta =
\left\{\begin{array}{ccl}
\frac{2\Gamma(\frac{i+1}{2})\Gamma(\frac{j+1}{2})}{\Gamma(\frac{i+j+2}{2})}
 & \mbox{if} &
i \;\mbox{and} \; j \; \mbox{even } , \; i,j  \in \mathbb{N} \cup \{ 0 \} \nonumber\\
0 & \mbox{if} &  i \mbox{ or } j \;\mbox{odd} , \nonumber\\
\end{array}\right.
\end{equation}
where $\Gamma$ is Euler's gamma function, it can be seen that the following recurrence holds:
$$
M_n(h) = \left\{ \begin{array}{lcc} M_{n-1}(h) & , & \mbox{if $n$ is even} \\ M_{n-2}(h) + \chi_n (\sqrt{h})^n& , & \mbox{if $n$ is odd}
\end{array} \right.
$$
where $M_1(h)=M_2(h)=\chi_1 \sqrt{h}$, and $\chi_n$ (with $n \geq 1$) is a real constant depending on $a_{ij}(c^2)$, $b_{ij}(c^2)$, $I_{ij}$ and $J_{ij}$, in all cases for those $i,j$ such that $i+j=n$. Consequently, $M_n(h)$ is an odd polynomial of $\sqrt{h}$, and therefore $h=0$ is always a root. Moreover, the remaining roots are distributed symmetrically around the origin. In particular, the maximum number of positive roots of $M_n(h)$ is (\ref{cotas}). Finally, we shall now prove that there exist suitable $P(x,y,z)$ and $Q(x,y,z)$ such that this maximum number is achieved for every $n$ because of the arbitrariness in the constants $\chi_n$, which in turn arises from the arbitrariness in the coefficients $a_{ij}(c^2)$ and $b_{ij}(c^2)$.  \bbox

\begin{remark}
{\rm Note that statement (iii) of Theorem \ref{teoeuler1} agrees with a classical result in the theory of limit cycles (see \cite{Li-Me-Pu}) that an $n$--degree polynomial perturbation of the harmonic oscillator $\dot{x} = -y + \epsilon (a x + P(x, y))$, $\dot{y} = x + \epsilon (a y + Q(x, y))$ has at most $(n-1)/2$ limit cycles if $n$ is odd and $(n-2)/2$ limit cycles if $n$ is even.}
\end{remark}

\begin{theorem}\label{teoeuler2}
Let us consider the Euler top (\ref{Euler1}) with $\alpha \beta < 0$, therefore having all the invariant spheres $\mathbb{S}_c$ ($c>0$) foliated with periodic orbits. Then, the following holds:
\begin{description}
\item[(i)] The most general analytic 1--parameter perturbation of (\ref{Euler1}) in $\mathbb{R}^3$ leaving invariant all the spheres $\mathbb{S}_c$ ($c>0$) is given by (\ref{Euler9}), with $A(x_1,x_2,x_3)$, $B(x_1,x_2,x_3)$ and $C(x_1,x_2,x_3)$ any analytic functions in $\mathbb{R}^3$ satisfying
\begin{equation}
\label{condicion-inv.casimir}
x_1 A(x_1,x_2,x_3) + x_2 B(x_1,x_2,x_3) + x_3 C(x_1,x_2,x_3) = 0 \ .
\end{equation}
In addition, a family of solutions of (\ref{condicion-inv.casimir}) is given by
\begin{equation}
\label{mnl}
A=x_3M-x_2N \: , \:\;\: B=x_1N-x_3L \: , \:\;\: C= x_2L-x_1M
\end{equation}
for arbitrary analytic functions $L(x_1,x_2,x_3)$, $M(x_1,x_2,x_3)$ and $N(x_1,x_2,x_3)$. Moreover, if the perturbation $(A, B, C)$ is polynomial and homogeneous of degree $m$, then the family (\ref{mnl}) provides the general solution of (\ref{condicion-inv.casimir}) for $L$, $M$ and $N$ some homogeneous polynomials of degree $m-1$.
\item[(ii)] Assume that {\small
\begin{equation} \label{Euler11-Benito}
I(h) = \oint_{H = h} \frac{A \left(x,y, \sqrt{c^2-(x^2+y^2)} \right) dy - B \left( x,y, \sqrt{c^2-(x^2+y^2)} \right) dx}{\sqrt{c^2-(x^2+y^2)}} \not\equiv 0 \ ,
\end{equation}}
where $H(x,y) = \frac{1}{2} (\alpha y^2 - \beta x^2)$. Then, for all $c>0$, the periodic orbits $\gamma_{h^*} \subset \mathbb{S}_c^{+}$ of (\ref{Euler1}) from which bifurcates a limit cycle of the perturbed system (\ref{Euler9}) satisfying (\ref{condicion-inv.casimir}) are given by
$$
\gamma_{h^*} = \{ (x_1, x_2, x_3) \in \mathbb{S}_c^{+} \ : \ \mathcal{H}(x_1, x_2, x_3) = c^2/(2 \mu_3) - h^* \} \ ,
$$
where $I(h^*)=0$, $\alpha \, h^* > 0$ and
$$
h^* < \frac{c^2}{2} \left( \frac{1}{\mu_3} - \max \left\{ -\alpha \: , \: \beta \right\} \right) \: , \:\;\: \mbox{or} \:\;\:
h^* > \frac{c^2}{2} \left( \frac{1}{\mu_3} - \min \left\{ -\alpha \: , \: \beta \right\} \right)
$$
if $\alpha > 0$ or $\alpha < 0$, respectively.

\item[(iii)] If $A(x_1,x_2,x_3)=x_3P(x_1,x_2,x_3^2)$ and $B(x_1,x_2,x_3)=x_3Q(x_1,x_2,x_3^2)$ verifying condition (\ref{condicion-inv.casimir}) with polynomials $P$ and $Q$ of maximum degree $n$, then a sharp upper bound of the maximum number of limit cycles of system (\ref{Euler9}) on $\mathbb{S}_c^{+}$ for any $c>0$ is $n-1$.
\end{description}
\end{theorem}
\noindent{\bf Proof.} Recalling that $\alpha+\beta+\gamma=0$ and imposing that the Casimir function $D(x_1,x_2,x_3) = x_1^2+x_2^2+x_3^2$ must be a first integral also for the perturbed system (\ref{Euler9}) for all $\epsilon$, we get that the functions $A(x_1,x_2,x_3)$, $B(x_1,x_2,x_3)$ and $C(x_1,x_2,x_3)$ must satisfy (\ref{condicion-inv.casimir}). In the particular case that the components of the perturbation field $(A, B, C)$ are homogeneous polynomials of degree $m$, Darboux showed \cite{Dar} that condition (\ref{condicion-inv.casimir}) is equivalent to the existence of homogeneous polynomials $L$, $M$ and $N$ of degree $m-1$ such that $A = z M - y N$, $B = x N - z L$ and $C = y L - x M$. Thus, statement (i) is proved.
\newline

Regarding statement (ii), we perform again the change of variables (\ref{mistetas}), which is a diffeomorphism in $\{(x_1,x_2,x_3) \in \mathbb{R}^3 : \: x_3 > 0 \}$. Recall that $\mathbb{S}_c^{+}$ is mapped into the disk $\Omega = \{ (x,y, c^2) \in \mathbb{R}^3 : \:  x^2+y^2 < c^2 \}$. Following analogous steps to those in the proof of Theorem \ref{teoeuler1}, the perturbed system (\ref{Euler9}), once reduced to the semispace $x_3 > 0$, restricted to the invariant disk $\Omega$ for any $c > 0$, and submitted to the time rescaling $d \tau = \sqrt{c^2-(x^2+y^2)} \; d t$, takes the form
\small{
\begin{equation}
\label{Euler-Pontryagin4}
\dot{x} = \frac{\partial H}{\partial y} + \epsilon \frac{A \left( x, y, \sqrt{c^2-(x^2+y^2)} \right)}{\sqrt{c^2-(x^2+y^2)}} \: , \:\;\:\;
\dot{y} = -\frac{\partial H}{\partial x} + \epsilon \frac{B \left(x, y, \sqrt{c^2-(x^2+y^2)} \right) }{\sqrt{c^2-(x^2+y^2)}}
\end{equation}}
with $H(x,y) = \frac{1}{2} (\alpha y^2 - \beta x^2)$. Note that the previous system is an analytic perturbation in the disk $\Omega$ of a Hamiltonian vector field.
\newline

Taking again into account that condition $\alpha \beta < 0$ implies that system (\ref{Euler-Pontryagin4}) with  $\epsilon = 0$ possesses on the invariant disk $\Omega$ a center at $(x,y)=(0,0)$ with period annulus foliated by the ellipses $\gamma_h = H^{-1}(h)$, by imposing the condition $\gamma_h \subset \Omega$ and taking into account Theorem \ref{Euler-Pontryagin3}, the same reasoning employed in the proof of Theorem \ref{teoeuler1} leads to the proof of statement (ii).
\newline

In order to prove statement (iii), we consider the case $A(x_1,x_2,x_3)=x_3P(x_1,x_2,x_3^2)$ and $B(x_1,x_2,x_3)=x_3Q(x_1,x_2,x_3^2)$, as indicated.
Assuming $\alpha > 0$ and $\beta < 0$ without loss of generality and parameterizing the ellipses $\gamma_h$ with $h > 0$ as in the proof of Theorem \ref{teoeuler1}, namely $x=\sqrt{-2 h /\beta} \cos\theta$ and $y=\sqrt{2 h /\alpha} \sin\theta$ with $\theta \in [0,2 \pi)$, we get
\begin{equation} \label{Euler113}
I(h) = \sqrt{h} \int_{0}^{2 \pi} ( \bar{P}(\theta, h, c) \cos\theta + \bar{Q}(\theta, h, c) \sin\theta )  d \theta \ ,
\end{equation}
where
\begin{eqnarray*}
\bar{P}(\theta, h, c) &:=& \sqrt{\frac{2}{\alpha}} \ P\left( \sqrt{\frac{-2 h }{ \beta}} \cos\theta, \sqrt{\frac{2 h }{\alpha}} \sin\theta, c^2 +2h \left[ \frac{\cos ^2 \theta}{\beta}-\frac{\sin ^2 \theta}{\alpha} \right] \right) \ , \\
\bar{Q}(\theta, h, c) &:=& \sqrt{\frac{-2}{\beta}} \ Q\left( \sqrt{\frac{-2 h }{\beta}} \cos\theta, \sqrt{\frac{2 h }{\alpha}} \sin\theta, c^2 +2h \left[ \frac{\cos ^2 \theta}{\beta}-\frac{\sin ^2 \theta}{\alpha} \right]\right) \ .
\end{eqnarray*}
Since $P$ and $Q$ are polynomials of degree $n$, that is, $P(x,y,z) = \sum_{i+j+k=0}^{n} a_{ijk} x^i y^j z^k$ and $Q(x,y,z) = \sum_{i+j+k=0}^{n} b_{ijk} x^i y^j z^k$, taking (\ref{integrales}) into account and using similar arguments to those in the proof of statement (iii) of Theorem \ref{teoeuler1}, it is straightforward to show that $I(h) = h M_{n-1}(h)$ with $M_{n-1}(h)$ a polynomial in the variable $h$ of degree at most $n-1$. In particular, the maximum number of positive roots of $M_{n-1}(h)$ is $n-1$, which can be achieved for suitable $P(x,y,z)$ and $Q(x,y,z)$. This completes the proof.  \bbox
\newline

A corollary of Theorem \ref{teoeuler2} for a class of homogeneous perturbations is the following:

\begin{corollary}\label{corol-Euler1}
Let us consider the Euler top (\ref{Euler1}) with $\alpha \beta < 0$, therefore having all the invariant spheres $\mathbb{S}_c$ ($c>0$) foliated with periodic orbits. Assume that the perturbation field $(A,B,C)$ in (\ref{Euler9}) is polynomial and homogeneous of degree $m$ satisfying (\ref{condicion-inv.casimir}) and of the form $A(x_1,x_2,x_3)=x_3P(x_1,x_2,x_3^2)$ and $B(x_1,x_2,x_3)=x_3Q(x_1,x_2,x_3^2)$. Then, the upper bound stated in Theorem \ref{teoeuler2} (iii) for the maximum number of limit cycles of system (\ref{Euler9}) on $\mathbb{S}_c^{+}$ for any $c>0$ is not achieved. More precisely, if $I(h)$ is the function (\ref{Euler11-Benito}), the following holds:
\begin{description}
\item[(i)] If $m=3$ or $m=5$, then $I(h) \equiv 0$.
\item[(ii)] If either $m=4$ or $m=6$, then $I(h) = h M_{k}(h)$ where $M_{k}$ is a polynomial of degree $k$ with $k=1$ or $k=2$, respectively. In particular, the maximum number of limit cycles of system (\ref{Euler9}) on $\mathbb{S}_c^{+}$ for any $c>0$ is $k$.
\item[(iii)] If $m=7$ and $I(h) \not\equiv 0$, then $I(h)$ has either one unique positive root $h^* > 0$ or none according to whether $\alpha \neq \beta$ or not, respectively. In addition, $h^* = \alpha \beta c^2 / (\beta-\alpha)$ does not depend on the perturbation field $(A,B,C)$.
\end{description}
\end{corollary}

\begin{example}
{\rm In case (iii) of Corollary \ref{corol-Euler1} we have $I(h) = \lambda \, h^2 [\alpha \beta c^2 + (\alpha-\beta) h]$ where the constant $\lambda = \pi (\lambda_1 \beta - \lambda_2 \alpha)/(-\alpha \beta)^{5/2}$ being $\lambda_1$ and $\lambda_2$ the coefficient of $P$ in $x_1 x_2^2 x_3$ and the coefficient of $Q$ in $x_1^2 x_2 x_3$, respectively. Thus, $I(h) \not\equiv 0$ if and only if $\lambda_1 \beta \neq \lambda_2 \alpha$. In this case, $I(h^*)=0$ where $h^* = \alpha \beta c^2 / (\beta-\alpha)$. Therefore when $\alpha > 0$ (hence $\beta <0$) and according to statement (ii) of Theorem \ref{teoeuler2}, in order to have a limit cycle of system (\ref{Euler9}) on $\mathbb{S}_c^{+}$ for any $c > 0$ the additional condition
\begin{equation}\label{eul-ben}
h^* < \frac{c^2}{2} \left( \frac{1}{\mu_3} - \max \left\{ -\alpha \: , \: \beta \right\} \right)
\end{equation}
must be satisfied. It is easy to see that always exists a $\mu_3 > 0$ such that (\ref{eul-ben}) is satisfied. More precisely, the choice of $\mu_3$ such that condition (\ref{eul-ben}) is fulfilled is as follows: if $\alpha + \beta > 0$ then take $0 < \mu_3 < (\beta - \alpha)/(\beta ( \alpha + \beta ))$ and when $\alpha + \beta < 0$ it must be $0 < \mu_3 < (\beta - \alpha)/(\alpha ( \alpha + \beta ))$. The symmetric case $\alpha + \beta = 0$ gives no restriction except $\mu_3 > 0$.
}
\end{example}

\begin{example}
{\rm In \cite{Puta}, the problem of stabilization of permanent rotations of the free rigid body with two controls about the intermediate principal axis is considered. In short, system (\ref{Euler9}) with $A(x_1,x_2,x_3) = - k x_3$, $B(x_1,x_2,x_3) = 0$ and $C(x_1,x_2,x_3) = k x_1$ is considered, where $k \in \mathbb{R}$ is the feedback gain parameter. Here we shall consider a modified version of this problem consisting of the perturbed field:
\begin{eqnarray*}
A(x_1,x_2,x_3) &=& -x_3 (k -x_1 + x_1 x_3^2) \ , \\
B(x_1,x_2,x_3) &=& x_2 x_3 (1+x_3^2) \ , \\
C(x_1,x_2,x_3) &=& k x_1 + x_1^2(x_3^2-1) - x_2^2 (1+x_3^2) \ .
\end{eqnarray*}
Notice that this perturbation satisfies the conditions of statement (iii) of Theorem \ref{teoeuler2} with $P(x_1,x_2,x_3) = -k + x_1(1-x_3)$ and $Q (x_1,x_2,x_3) = x_2(1+x_3)$ and therefore at most 1 limit cycle of system (\ref{Euler9}) can appear on each $\mathbb{S}_c^{+}$ for any $c>0$. Direct computations show that the function $I(h)$ of (\ref{Euler11-Benito}) is $I(h) = \lambda \, h [-2 \alpha \beta + (\alpha+\beta) h]$ where $\lambda$ is a non--vanishing constant. Thus, $I(h^*)=0$ where $h^* = 2 \alpha \beta / (\alpha + \beta)$. Taking $\alpha > 0$ (hence $\beta <0$) and using statement (ii) of Theorem \ref{teoeuler2}, the additional condition (\ref{eul-ben}) must be satisfied to have a limit cycle of system (\ref{Euler9}) on $\mathbb{S}_c^{+}$. It is easy to show that (\ref{eul-ben}) is always satisfied for any $c>0$ and $\mu_3 >0$ under the condition $\alpha + \beta > 0$. On the contrary, when $\alpha + \beta < 0$ two cases arise fulfilling restriction (\ref{eul-ben}) according to whether $c$ is greater or not than $c^* := 2 \sqrt{\beta / (\alpha+\beta)}$. More precisely, when $\alpha + \beta < 0$, either $0 < c < c^*$ and $0 < \mu_3 < (\alpha+\beta) c^2 / [4 \alpha \beta - \alpha (\alpha+\beta) c^2]$ or $c \geq c^*$ and $\mu_3 >0$.
}
\end{example}

\mbox{}

\noindent {\bf Acknowledgments.}

\noindent The first author (I.G.) is partially supported by a MCYT/FEDER grant number
MTM2008-00694 and by a CIRIT grant number 2009 SGR 381. The second author (B.H.-B.) would
like to acknowledge the kind hospitality at Lleida University during which part of this
work was developed.

\pagebreak

\end{document}